\documentclass{aa}
\usepackage{graphicx}
\usepackage{txfonts}
\newcommand{\lund}{Lund Observatory, Division of Astrophysics, Department of Physics, Lund University, Box 43, SE-221 00 Lund, Sweden}

\newcommand{\amnh}{Department of Astrophysics, American Museum of Natural History, 200 Central Park West, New York, NY 10024, USA}
\newcommand{\strasbourg}{University of Strasbourg Institute for Advanced Study, 5 all\'ee du G\'en\'eral Rouvillois, F-67083 Strasbourg, France}
\newcommand{\columbia}{Department of Astronomy, Columbia University, New York, NY 10027, USA}
\newcommand{\tsinghua}{Department of Astronomy, Tsinghua University, Beijing 100084, China}

\begin{document}

   \title{Pre-supernova feedback sets the star cluster mass function to a power law and reduces the cluster formation efficiency}

   \subtitle{}

   \author{Eric~P.~Andersson
          \inst{1}
          \and
          Mordecai-Mark~Mac~Low
          \inst{1}
          \and
          Oscar~Agertz
          \inst{2}
          \and
          Florent~Renaud
          \inst{2,3}
          \and
          Hui~Li
          \inst{4,5}
          }

   \institute{\amnh\\
              \email{eandersson@amnh.org, mordecai@amnh.org}
         \and
             \lund\\
              \email{oscar.agertz@astro.lu.se, florent.renaud@astro.lu.se}
         \and
             \strasbourg
         \and
             \tsinghua\\
             \email{hliastro@tsinghua.edu.cn}
         \and
             \columbia
             }
   \date{Received Month day, year; accepted Month day, year}

  \abstract
   {The star cluster initial mass function is observed to have an inverse power law exponent around $2$, yet there is no consensus on what determines this distribution, and why some variation is observed in different galaxies. Furthermore, the cluster formation efficiency covers a range of values, particularly when considering different environments. These clusters are often used to empirically constrain star formation and as fundamental units for stellar feedback models. Detailed galaxy models must therefore accurately capture the basic properties of observed clusters to be considered predictive.}
   {Studying how feedback mechanisms acting on different timescales and with different energy budgets affect the star cluster mass function and cluster formation efficiency.}
   {We use hydrodynamical simulations of a dwarf galaxy as a laboratory to study star cluster formation. We test different combinations of stellar feedback mechanisms, including stellar winds, ionizing radiation, and supernovae.}
   {Each feedback mechanism affects the cluster formation efficiency and cluster mass function. Increasing the feedback budget by combining the different types of feedback decreases the cluster formation efficiency by reducing the number of massive clusters. Ionizing radiation is found to be especially influential. This effect depends on the timing of feedback initiation, as shown by comparing early and late feedback. Early feedback occurs from ionizing radiation and stellar winds with onset immediately after a massive star is formed. Late feedback occurs when energy injection only starts after the main-sequence lifetime of the most massive SN progenitor, a timing that is further influenced by the choice of the most massive SN progenitor. Late feedback alone results in a broad, flat mass function, approaching a log-normal shape in the complete absence of feedback. Early feedback, on the other hand, produces a power-law cluster mass function with lower formation efficiency, albeit with a steeper slope than that usually observed.}
   {}
   \titlerunning{Star cluster mass function and CFE}
   \authorrunning{Eric~P.~Andersson et al.}
   \maketitle
%

\section{Introduction}
Star clusters represent an opportunity to constrain star formation and stellar feedback models used to understand galaxies, a notoriously challenging problem due to its physical complexity and wide dynamical range \citep[][]{Somerville&Dave2015,Naab&Ostriker2017}.
Properties of clusters, such as their mass function, have been a subject for observations for a long time in our own Galaxy \citep[see][]{Lada&Lada2003}, in our galactic neighbors the Magellanic Clouds \citep[e.g.,][]{Elmegreen&Efremov1997,Hunter+2003,Larsen2009} and M31 \citep[e.g.,][]{Vansevicius+2009,Johnson+2015,Johnson+2017}, as well as many other galaxies \citep[e.g.,][]{Zhang&Fall1999,Larsen2009,Cook+2012,Adamo+2015,Chandar+2014,Chandar+2016,Adamo+2020a,Adamo+2020b,Wainer+2022,Cook+2023}, providing strong empirical constraints on models. 

Typically, the observed mass function for young clusters follows an inverse power law, possibly with a truncation at higher masses \citep[see][]{PortegiesZwart&McMillan&Gieles2010,Krumholz&McKee&Bland-Hawthorn2019,Adamo+2020a}, although the truncation is debated \citep[][]{Mok+2019}.
The truncation mass varies between different types of galaxies, while the slope $\alpha$ ranges from $-1.5$ to $-2.5$ \citep[see figure 10 in][]{Adamo+2020a}.
The slope $\alpha\approx-2$ is somewhat unsurprising, since this is the slope of a scale-free distribution \citep[][]{Guszejnov+2018}, which results from gas fragmentation under gravity and turbulence \citep[][]{Elmegreen&Efremov1997,Elmegreen2002}.
The slope of the distribution is similar for molecular clouds \citep[][]{Colombo+2019,Mok+2020,Rosolowsky+2021}, indicating that there could be a direct mapping between the two, though this would imply that the star formation efficiency and timescale are independent of the cloud mass \citep{Clark+2007}.
Nonetheless, understanding the observed variations remains an unsolved problem, particularly their dependence on the environment.

Several numerical studies address the mechanisms that determine the shape of the cluster mass function, primarily focusing on sub-grid modeling of star formation and the effects of prescribed feedback.
By treating star formation in units of star clusters in cosmological simulations, \citet{Li+2017} found that $100\%$ star formation efficiency per free-fall time (SFE) resulted in a log-normal cluster mass function peaking around $10^4\ {\rm M}_\sun$, while models with lower SFE resulted in a Schechter-like mass function with $\alpha\sim 2$ and a cutoff mass with a dependence on the star formation rate (SFR) or SFR surface density \citep{Li&Gnedin&Gnedin2018}.
Interestingly, a log-normal cluster mass distribution is often found in simulations of galaxies undergoing major mergers \citep[][]{Renaud+2015,Maji+2017}, although exceptions are also reported \citep[][]{Li+2022}.
Major mergers tend to increase the amount of dense gas, which accelerates star formation \citep[i.e. shorter depletion time,][]{Renaud+2014,Li+2022,SegoviaOtero+2022}.
Similarly to high SFE, this could provide the environment necessary for
massive cluster formation \citep[see, e.g.,][]{Kravtsov&Gnedin2005,Parmentier&Gilmore2007,Renaud+2015}.
Increasing the feedback strength by boosting the momentum input from supernovae (SNe) by a factor of five (in order to match global SFR with empirical data) resulted in increased variation between different choices of SFE, with higher SFE resulting in mass function extending toward higher masses, shallower slopes ($\alpha\approx2.5$) at their upper end, and a log-normal distribution emerged from high values of SFE \citep[$200\%$,][]{Li&Gnedin&Gnedin2018}.
Furthermore, \citet{Li&Gnedin&Gnedin2018} found that reducing the momentum boost factor to three significantly steepened the slope of the mass function for young clusters.
It is clear that feedback impacts the cluster mass function; however, there is no systematic study of the impact of different feedback sources in these studies.

Another open question is what fraction of stars emerge in bound clusters.
Although most stars originate in groups and sub-clusters that form in gas concentration generated by turbulent motions inside giant molecular clouds \citep[GMCs, see, e.g.,][]{MacLow&Klessen2004,Gomez&Vazquez-Semadeni2014,Krumholz+2014} and assemble into clusters \citep[see, e.g.,][]{Wall+2019,Wall+2020,Cournoyer-Cloutier+2023}, the vast majority are rapidly dissolved due to gas ejection \citep[see seminal work by][]{Lada&Lada2003}.
The cluster formation efficiency (CFE), quantifying the fraction of newly born stars in clusters has been estimated for a large number of observed galaxies \citep[see, e.g.,][]{Goddard+2010,Adamo+2011,Annibali+2011,Adamo+2015,Pasquali+2011,Cook+2012,Lim&Lee2015,Hollyhead+2016,Chandar+2017,Ginsburg+2018,Messa+2018,Fensch+2019,Adamo+2020b,Chandar+2023,Cook+2023}.
These observations are typically interpreted as a function of star formation intensity, as measured by star formation rate density $\Sigma_{\rm SFR}$ to compare different environments.
Note that CFE measurements depend on the cluster age \citep[][]{Adamo+2020a}, making the interpretation sensitive to observational tracers \citep[][]{Chandar+2017,Chandar+2023}. 
For young clusters (<~$10$~Myr), CFE is typically found to be high (at least a few tens of percent) at $\Sigma_{\rm SFR}>0.1\ {\rm M}_\sun\, {\rm yr}^{-1}\, {\rm kpc}^{-2}$ \citep[see, e.g.,][]{Adamo+2011,Adamo+2020b}, while at lower $\Sigma_{\rm SFR}$, CFE scatters from a few percent up to values similar to those measured at high $\Sigma_{\rm SFR}$ \citep[][]{Goddard+2010,Lim&Lee2015,Chandar+2017}. 

Theoretical predictions for the CFE have received similar attention to the mass function.
\citet{Kruijssen2012} argued that a scaling between CFE and $\Sigma_{\rm SFR}$ arises due to a higher bound fraction for more intense star formation at higher gas densities, with saturation in CFE of around $70\%$ at the highest $\Sigma_{\rm SFR}$.
As in \citet{Kruijssen2012}, an increasing CFE with respect to $\Sigma_{\rm SFR}$ was also found in models by \citet{Li+2017,Li&Gnedin&Gnedin2018}.
In this case, a choice of higher SFE shifted the normalization of the scaling toward higher CFE, best matching observations at SFE $\gtrsim50\%$.
Surprisingly, \citet{Li&Gnedin&Gnedin2018} found that boosting feedback only marginally changed the normalization of the CFE-$\Sigma_{\rm SFR}$ relation.
Furthermore, they found that at high-$\Sigma_{\rm SFR}$, the saturation of CFE was related to the maximum cluster mass rather than tidal effects.

Clustered star formation has received attention for star formation and stellar feedback modeling of galaxies, both for simulations in isolation and cosmological environments \citep[][]{Naab&Ostriker2017}.
For example, numerical models of stellar feedback injection are sensitive to clustered SN explosions \citep[][]{Keller+2014,Agertz&Kravtsov2015,Smith+2021} and runaway stars originating from clusters \citep[][]{Ceverino&Klypin2009,Andersson+2020,Steinwandel+2022,Andersson+2023}.
Furthermore, these models can provide predictions for the initial cluster mass function and CFE if measured before the cluster experiences significant dynamical evolution.
In simulations resolving individual stars, \citet{Hislop+2021} found a variation in the mass function and CFE for bound clusters for different choices of SFE, raising concerns about the ability of star formation recipes with assumed SFE to accurately model cluster formation.
In addition, \citet{Hislop+2021} found that photoionizing radiation reduced the CFE by almost $40\%$ for $2\%$ SFE.
With a similar model assuming $100\%$ SFE, \citet{Lahen+2023} found that the CFE is scattered between $1$ and $100\%$ between different outputs in a dwarf galaxy with low $\Sigma_{\rm SFR}$.
This scatter remains unexplained, and to the best of our knowledge, no systematic investigation of how different sources of feedback affect these properties has been conducted, particularly concerning stellar winds.

To this end, we investigate hydrodynamical simulations of dwarf galaxies and systematically compare how the cluster population is affected by different combinations of stellar feedback.
We include core-collapse and type Ia SNe, ionizing radiation using a radiative transfer model, and stellar winds.
Furthermore, we test the effect of varying the maximum progenitor mass of core-collapse SNe \citep[][]{Janka2012}. 

We describe the model and simulation setup in Section~\ref{sec:Method} and present our results in Section~\ref{sec:Results}. We discuss these results in a broader context and test model assumptions in Section~\ref{sec:Discussion}. Finally, Section~\ref{sec:Summary} summarizes our work.

\section{Numerical method and initial conditions}\label{sec:Method}

The suite of galaxies presented are simulated using {\sc ramses-rt} \citep[][]{Roshahl+2013,Rosdahl&Teyssier2015}, a multi-group radiative transfer extension of the adaptive mesh refinement and $N$-body code {\sc ramses} \citep[][]{Teyssier2002}. 
{\sc ramses} has an oct-tree hierarchical grid providing adaptive resolution and solves the fluid equations using a second-order unsplit Godunov method with the HLLC approximate Riemann solver \citep[][]{Toro+1994}. 
To avoid spurious oscillations, we apply a MinMod slope limiter to reconstruct the piecewise linear solution for the Godunov solver \citep[see, e.g.,][]{Toro2009}.
A monatomic, ideal gas equation of state with an adiabatic index $\gamma = 5/3$ is applied to close the set of fluid equations.
Radiation in {\sc ramses-rt} is treated via the momentum-based radiative transfer equations with an M1 closure relation for the Eddington tensor \citep[][]{Levermore1984}.
Our simulations account for non-equilibrium gas chemistry and cooling, tracking the ionization fractions of \ion{H}{I}, \ion{H}{II}, \ion{He}{II}, and \ion{He}{III} by advecting passive scalars on the grid \citep[see][for details]{Roshahl+2013,Rosdahl&Teyssier2015}.
Furthermore, the code models the non-equilibrium evolution of molecular hydrogen (H$_2$) following \citet{Nickerson+2018}.
All thermochemistry evolves by a semi-implicit method, described in \citet[][]{Roshahl+2013}.

Stars and dark matter are tracked by particles, with a mass resolution of $100\ {\rm M}_\sun$ and $10^4\ {\rm M}_\sun$, respectively.
For gravity, the Poisson equation is solved on the grid using the multi-grid method \citep[][]{Guillet&Teyssier2011}, with particle mapping via the cloud-in-cell method \citep[][]{Hockney&Eastwood1988}.
The refinement strategy employs a quasi-Lagrangian technique (aiming for eight particles in each cell) and cell refinement at a mass threshold of $800\ {\rm M}_\sun$.
The refinement is limited to 14 levels, with a cell size of $3.6$~pc on the maximum refinement level.

The initial conditions mimic an isolated dwarf galaxy using the tool {\sc MakeDiscGalaxy} \citep{Springel2005}.
These include a dark matter halo following a NFW-profile \citep[][]{NFW1997} with virial mass $M_{\rm vir}=10^{10}\ {\rm M}_\sun$, spin parameter $\lambda=0.04$, and concentration parameter $c=15$. 
Embedded within this halo is a gas ($M_{\rm g,disc}=7\times10^{7}\ {\rm M}_\sun$) and stellar ($M_{\rm s, disc}=10^{7}\ {\rm M}_\sun$) disc with an exponential radial density profile with scale length $1.1$~kpc.
The gas has an initial temperature of $10^4$~K and vertical distribution set by hydrostatic equilibrium, while stars have a vertical distribution of a Gaussian with a scale height of $0.7$~kpc.
The initial metallicity of the gas is set to $0.1\ {\rm Z}_\sun$.
These initial conditions were studied using a different model in \citet{Andersson+2023}, and are highly similar to those studied in \citet{Smith+2021}.

\subsection{Star formation and stellar feedback}\label{sec:sf_and_fb}

\begin{figure*}
    \centering
    \includegraphics[]{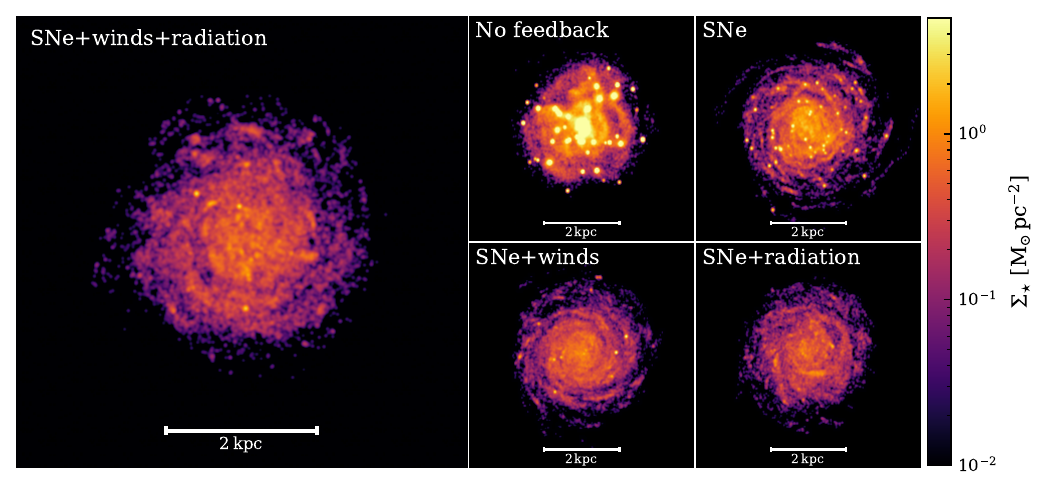}
    \caption{Face on view of the stellar surface density of all simulations. Each panel shows the last snapshot of the simulation ($t=1$~Gyr). There is a notable increase in the number of visible star clusters in simulations with less feedback, compare, e.g., SNe and SNe+wind+radiation.}
    \label{fig:surface_density}
\end{figure*}

Star formation proceeds by stochastically forming star particles (each with mass $100\ {\rm M}_\sun$) in all cells with gas density $\rho>10^3\ {\rm cm}^{-3}$ and temperature $<10^4$~K at each time step.
The stellar mass formed is removed from the gas mass of that cell to ensure mass conservation.
In each cell fulfilling the star formation criterion, the number of star particles to form is determined by sampling a Poisson distribution parameterized by $m_{\star}$ per $100\ {\rm M}_\sun$, where 
\begin{equation} 
m_{\star} = \dot{\rho}_{\rm sf} V \Delta t,
\end{equation} the cell volume is $V$, and the timestep $\Delta t$. A Schmidt-like recipe is used for the local volumetric star formation rate density 
\begin{equation}\label{eqn:sfr}
    \dot{\rho}_{\rm sf} = \epsilon_{\rm ff}\frac{\rho}{t_{\rm ff}},
\end{equation}
where $\epsilon_{\rm ff}=0.1$ is the SFE \citep[see, e.g.,][for discussion regarding this choice]{Grisdale+2019}, and $t_{\rm ff}=(3\pi/[32{\rm G}\rho])^{1/2}$ is the local gas free-fall time, where ${\rm G}$ is the gravitational constant.

Each star particle inherits the oxygen and iron abundance of its natal gas to track metallicity $M_{\rm Z}=2.09M_{\rm O}+1.06M_{\rm Fe}$  (based on the Solar mixture, see \citealt{Asplund+2009}, for details see, e.g., \citealt{Kim+2014,Agertz+2021}). 
Elements are tracked by advecting passive scalars on the grid and primarily evolve via enrichment linked to stellar feedback. 
We use the yield data for stellar winds and core-collapse SNe from {\sc NuGrid} \citep[][]{Pignatari+2016,Ritter+2018}, and for SN Ia from \citet{Seitenzahl+2013}.

The stellar feedback model employs the method described in \citet{Agertz+2020} to model stellar winds and SNe (core-collapse and type Ia). Energy injection from radiation is computed by the radiative transfer solver.
Stellar winds and SNe inject the appropriate combinations of mass, metals, momentum, and thermal energy.
These quantities are injected equally between eight cells \citep[see][for details]{Agertz+2013,Agertz&Kravtsov2015}. 
We assume the initial mass function from \citet{Kroupa2001} for all feedback processes.

Fast stellar winds injected by massive ($>8\ {\rm M}_\sun$) stars have mass-loss rates computed using functions fitted to {\sc starburst99} \citep[][]{Leitherer+1999} evolutionary tracks of single-age stellar populations \citep[see][for a presentation of these fits]{Agertz+2013}.
These winds act as a source of momentum with a wind velocity of $1000\ {\rm km}\, {\rm s}^{-1}$.
Our model also considers mass loss from AGB stars as a source of chemical enrichment (injected without any wind velocity).
The AGB mass loss rate calculation uses an IMF-averaged initial-final mass relation \citep[][]{Kalirai+2008} with a mass range 0.5--8$\ {\rm M}_\sun$ as the AGB progenitor stellar mass.

Core-collapse SNe are stochastically sampled as discrete events using the main sequence lifetimes from \citet{Raiteri+1996}. For the primary suite of simulations, we assume progenitors in the mass range 8--30$\ {\rm M}_\sun$ but test a higher limit ($100\ {\rm M}_\sun$) in one simulation.
The conservative choice is motivated by the uncertainty of the upper limit due to the direct collapse into a black hole (see Section~\ref{sec:Discussion} for a detailed discussion).
In practice, we assume that the most massive stars ($>30\ {\rm M}_\sun$) undergo direct collapse into a black hole without the injection of mass, momentum, or energy unless otherwise stated.
In the event of a SN, $10^{51}$~erg of energy is injected into the gas surrounding its location.
Furthermore, core-collapse SNe are assumed to release momentum equivalent to $12\ {\rm M}_\sun$ at a velocity of $3000\ {\rm km}\, {\rm s}^{-1}$, calibrated to {\sc starburst99}.
For explosions where the cooling radius of the SN blast-wave is unresolved by more than 6 cells, the fraction of the energy expected to result in the terminal momentum of the explosion is injected as gas momentum rather than thermal energy following \citet{Kim&Ostriker2015}. 
This model assumes a terminal momentum of $2.95\times10^5\ {\rm M}_\sun\, {\rm km}\, {\rm s}^{-1}$.
Mass loss for core-collapse SNe is computed from the fit by \citet{Woosley&Weaver1995}.
Our model accounts for type Ia SNe via the time-delay distribution following \citet{Maoz&Graur2017}, assuming a time delay of $38$~Myr and rate of $2.6\times10^{-13}\ {\rm yr}^{-1}\ {\rm M}_{\sun}^{-1}$ to normalize the distribution.
For type Ia SNe, we assume a mass loss of $1.4\ {\rm M}_\sun$ \citep[][]{Chandrasekhar+1931} and the momentum release equivalent to $10^{51}$~erg.

To incorporate radiation feedback, we apply the evolving spectral energy distribution from \citet{Bruzual&Charlot2003} to each stellar particle, employing six separate photon groups (see Table 1 in \citealt{Agertz+2020}).
Briefly described, these groups are selected to incorporate scattered infrared photons (non-thermal), direct radiation pressure from optical and far-ultraviolet, and Lyman-Werner radiation for H$_2$ dissociation.
The remaining three groups account for photo-ionizing radiation around the ionizing energies of \ion{H}{I}, \ion{He}{I}, and \ion{He}{II}.
Production of infrared photons from absorption of optical and ultraviolet photons by dust is included following the methods described in \citet{Rosdahl&Teyssier2015} and \citet{Kimm+2017}, assuming a dust opacity of $10\,Z\,{\rm cm}^{2}\,{\rm g}^{-1}$ for infrared photons and $1000\,Z\,{\rm cm}^{2}\,{\rm g}^{-1}$ for photons of higher energy, where $Z$ is the gas metallicity in units of the solar value $Z_\sun=0.02$ \citep[see][for details]{Agertz+2020}.

\section{Results}\label{sec:Results}

Throughout this section, we investigate the properties of young star clusters as a population.
Unless otherwise stated, we search the outputs for clusters using a friends-of-friends (FOF) algorithm on the positions of stars younger than $25$~Myr with a linking length of $4$~pc.
The FOF groups are refined by removing stars that are not gravitationally bound\footnote{We define particles to be gravitationally bound to the cluster if their mutual specific gravitational energy ($e_{\rm pot,i}=\sum_{j\neq i} {\rm G}m_j/|r_j-r_i|$) exceeds their specific kinetic energy.} to the group and considering the groups with at least five remaining members as clusters.
We show how our main result depends on different choices of these parameters in Appendix~\ref{sec:fof_parameters}. 

We present simulations executed from the same initial conditions with different types of feedback activated (see Section~\ref{sec:Method}). 
Each simulation is labeled by what feedback mechanisms are active during run time, with five different combinations: 1) no feedback activated; 2) SNe only; 3) SNe and stellar winds; 4) SNe and radiation; 5) SNe, stellar winds and radiation (see, e.g., Figure~\ref{fig:surface_density} for an example of the labels).
The main suite of simulations runs for $1000$~Myr with an output cadence of $25$~Myr.
Furthermore, we omit outputs in the first $300$~Myr to avoid spurious effects from the initial conditions.

Figure~\ref{fig:surface_density} shows the stellar surface density of the final output in each simulation.
In this figure, it is apparent that introducing more feedback decreases the number of visible clusters.
This result is persistent throughout the simulation, and, as we will show later, this is primarily the case for more massive star clusters.

\subsection{Cluster formation efficiency}\label{sec:cfe}

\begin{figure}
    \centering
    \includegraphics[width=0.45\textwidth]{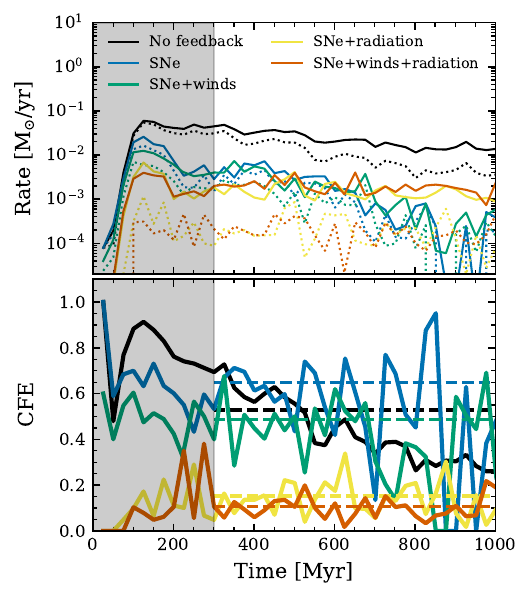}
    \caption{{\it Top:} SFR (filled line) and CFR (dotted line) as a function of time for all simulations. Both these quantities are measured on the same timescale ($25$~Myr). {\it Bottom:} The CFE as a function of time for all simulations. Dashed horizontal lines indicate the fractions of cluster mass to total stellar mass formed after $300$~Myr (outside gray shaded region).}
    \label{fig:cfr}
\end{figure}

The upper panel of Figure~\ref{fig:cfr} shows the SFR and cluster formation rate (CFR), while the lower panel shows their ratio ($\rm CFE = SFR/CFR$) to demonstrate how simulations with more feedback result in galaxies with fewer clusters.
We measure both these quantities on the same timescale ($25$~Myr).
For simulations including feedback, the CFE remains roughly constant in each model, albeit with a large short-term variation.
With radiation feedback, the mean CFE is $\sim15\%$, in stark contrast to the simulations with only SNe and SNe+winds, where the mean CFE is $\sim65\%$ and $\sim55\%$, respectively.
Without radiation, stellar winds suppress the CFE compared to only SNe. However, radiation dominates over stellar winds, resulting in similar CFE in the two simulations with radiation and with or without stellar winds. 
In the case of no feedback, the CFE steadily decreases with time, possibly as a result of the rapid decrease in gas fraction (the mass fraction of cold gas decreases by $40\%$, in contrast to the simulations with feedback where it remains above $80\%$ throughout the entire simulation). 
The absence of feedback results in a clumpy morphology as expected for the initially high gas fraction \citep{Li+2005,Renaud+2021}. Furthermore, a higher gas fraction can result in the formation of more clusters \citep[][]{Li+2005}.

\begin{figure}
    \centering
    \includegraphics[width=0.45\textwidth]{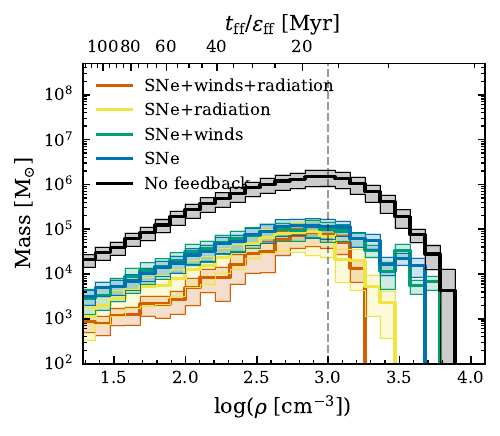}
    \caption{Average mass distribution of gas densities inside clusters. The distribution is the mean of the mass-weighted histograms of gas densities inside clusters. The inside of a cluster is the spherical region enclosing all cluster members. The shaded region indicates the standard deviation around the mean value. The dashed vertical line indicates the density threshold we use for star formation. The top horizontal axis shows the ratio between the free-fall time and $\epsilon_{\rm ff}=0.1$ (see Section~\ref{sec:sf_and_fb} for more details).}
    \label{fig:dmdrho}
\end{figure}

\begin{figure*}
    \centering
    \includegraphics[width=0.85\textwidth,trim={0 0 0 0.0cm},clip]{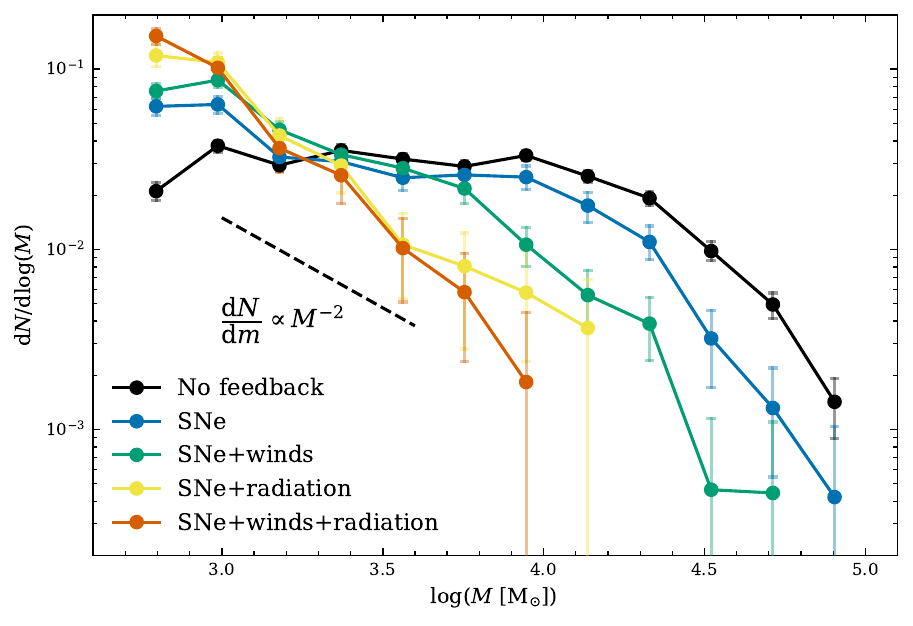}
    \caption{Cluster initial mass function, normalized to the total number of clusters (defined as objects with ages $<25$~Myr) found in each simulation after $300$~Myr. The error bars indicate the $16$\textsuperscript{th} and $84$\textsuperscript{th} percentiles computed using bootstrapping. The dashed line shows the slope of scale-free formation models, often associated with the initial cluster mass function (see text for details).}
    \label{fig:mass_function}
\end{figure*}

The immediacy of radiation and stellar winds, which begin directly after star formation in our model, implies that these processes immediately regulate local star formation,\footnote{By local, we mean the in-situ star formation of the cluster in contrast to accreted stars formed elsewhere.} in contrast to the simulation with only SNe where the onset of feedback after star formation is delayed by $\sim6$~Myr (the lifetime of a $30\ {\rm M}_\sun$ progenitor).
Note that feedback affects gas on kiloparsec scales; however, complexity in the spatial coupling of the different feedback sources \citep[e.g.][]{Ohlin+2019}
makes it challenging to quantify how clusters affect each other. We leave this to future work.

Although stellar winds also act promptly, they impart less energy and momentum than radiation, explaining the lower CFE in the simulation with SNe+winds+radiation compared to the one with only SNe+winds.
We return to this notion in Section~\ref{sec:formation_time} when we investigate the cluster formation time.

Feedback from stars affects the density structure of the star-forming gas, as shown in Figure~\ref{fig:dmdrho}. 
This figure shows the mass-weighted gas density distribution for cells inside clusters, calculated by the mean distribution of all clusters in a given simulation.
The top axis of this figure show the cell-based depletion time of star-forming gas, defined as gas density divided by star formation rate density, here shown as free-fall time per $\epsilon_{\rm ff}$ (see Equation~\ref{eqn:sfr}).
Absent radiation, stellar winds, or feedback from more distant SNe, only local thermal and turbulent forces prevent the gas from collapsing.
In this case, dense gas is rapidly consumed ($10$~Myr at $\rho\approx10^{3.5}\ {\rm cm}^{-3}$) before the first local SNe.
In simulations with radiation, we note a sharp drop in gas mass with increasing density above $\sim1000\ {\rm cm}^{-3}$ compared to simulations without radiation.
This transition is at the density threshold applied for star formation (see Section~\ref{sec:sf_and_fb}).
We also find that stellar winds play a smaller role in shifting the distribution toward lower densities.
Notably, the highest densities have depletion times on the order of the timescale for core-collapse SNe, although additional gas is likely accreted from the surrounding material.
The increase in $\dot{\rho}_{\star}$ at higher gas densities in simulations without radiation likely plays a role similar to changing the SFE, which has been shown to result in higher CFE \citep[see, e.g.,][]{Li&Gnedin&Gnedin2018,Hislop+2021}.
Furthermore, this can explain why \citet{Hislop+2021} found that including ionizing radiation and thus shifting the density distribution to lower peak values resulted in a $\sim30\%$ decrease in CFE.
We emphasize that the details of the simplified star formation models used in galaxy simulations must be sensitive to numerical resolution, in particular, the choice of star formation density threshold and SFE.

\subsection{Cluster initial mass function}\label{sec:mass_function}

We show the initial cluster mass function of each simulation in Figure~\ref{fig:mass_function}, derived by summing all identified clusters, which by assumption have ages $<25$~Myr.
When all feedback sources are active, the mass function scales as a power-law with a slope $\alpha\approx-3$, slightly steeper than the canonical mass function \citep[see, e.g.,][]{PortegiesZwart&McMillan&Gieles2010,Krumholz&McKee&Bland-Hawthorn2019}.
However, the cluster mass function in dwarf galaxies is often observed to have a steeper slope \citep[][]{Adamo+2020a}, although not to the extent found here.
Removing sources of feedback shifts the power law section of the mass function toward higher masses, leaving a plateau at lower masses if radiation is turned off. If radiation is included, there is no plateau.
Note that our simulations produce only a few massive clusters (particularly for low CFE).
We show the robustness of the derived mass function with error bars indicating $16$\textsuperscript{th} and $84$\textsuperscript{th} percentiles, each computed using bootstrapping.

\begin{figure}
    \centering
    \includegraphics[width=0.45\textwidth]{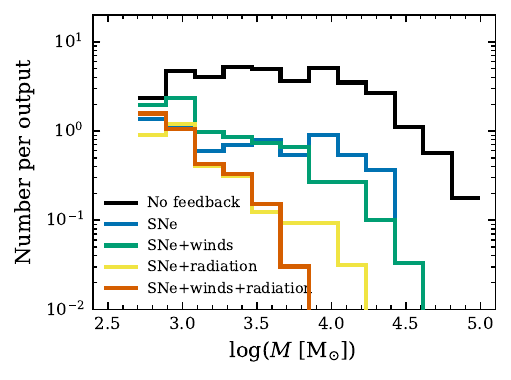}
    \caption{Histogram showing the cluster mass function, normalized such that the $y$-axis gives the number of clusters expected when looking at a random simulation snapshot. Notably, all simulations have roughly the same number of low-mass clusters. This figure is identical to Figure~\ref{fig:mass_function} except each line is normalized differently.}
    \label{fig:mass_histrogram}
\end{figure}

Figure~\ref{fig:mass_histrogram} shows the mass function normalized to the number of clusters in a given snapshot, i.e., the average number of young ($<25$~Myr) clusters found at any given time. 
While the number of high-mass clusters differs significantly, there are a similar number of low-mass clusters in all simulations: the average number of clusters per output shows a factor two difference at most. 
We conclude that the lower value for CFE in simulations with more feedback sources is due to an absence of the most massive clusters.

\begin{figure}
    \centering
    \includegraphics[width=0.45\textwidth]{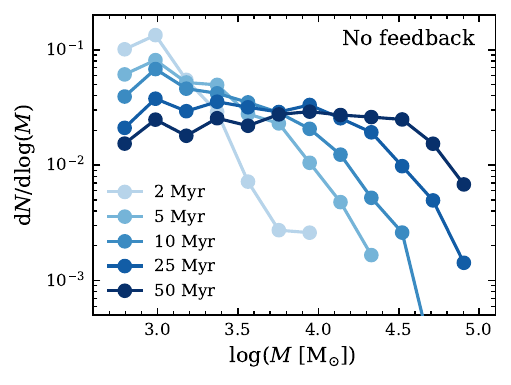}
    \caption{Cluster mass function for the simulation without feedback applying different age limits to the cluster finding algorithm. The age limit serves as a way to reduce the cluster formation time in a simulation where no stellar feedback is responsible for such limitations. Notably, this shifts the mass function to resemble simulations with feedback.}
    \label{fig:nofb_dndm}
\end{figure}

In the previous section, we noted that the early onset of local star formation regulation from pre-SN feedback affects the formation of star clusters.
To test how the mass function is affected by the limitation of the growth time rather than the growth rate, we employ the cluster-finding algorithm with different age limits to the simulation without any feedback. 
The result is in Figure~\ref{fig:nofb_dndm}, indicating a truncation at high masses similar to that set by introducing feedback. 
We show a similar figure for all other simulations in Appendix~\ref{sec:fof_parameters}.
The distribution is steeper at lower mass for age limits $<10$~Myr and is most similar to the slopes of the complete feedback model between 2--5~Myr.
We find a similar effect in simulations with delayed feedback (only SNe), shown in Appendix~\ref{sec:fof_parameters}. 
Since the slope at the upper end of the mass function is similar for all age limits, each cluster in a population must grow exponentially.
The rate at which gas is converted into stars depends on the gas mass (see Equation~\ref{eqn:sfr}); therefore, exponential growth is expected.
There is a complexity in this behavior, and we discuss this further in the next Section.

\subsection{Cluster formation time}\label{sec:formation_time}

\begin{figure*}
    \centering
    \includegraphics[width=1.0\textwidth]{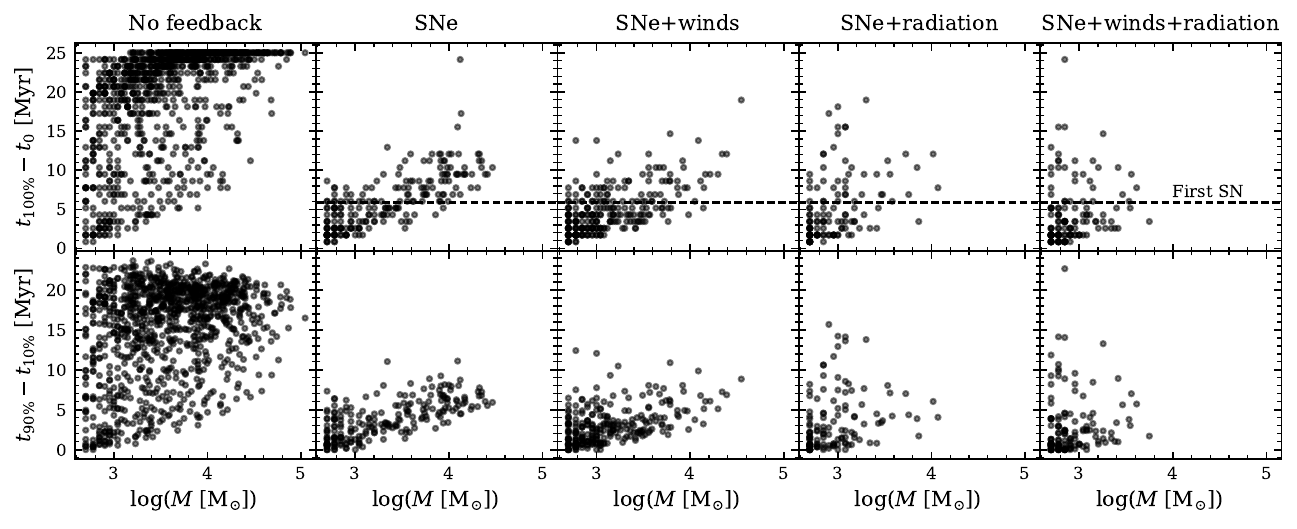}
    \caption{The formation time of individual clusters as a function of their mass for each simulation. The top row shows the time between when the first star was born $t_0$ and when the last star formed $t_{100\%}$, while the bottom row shows the time between when $10$ and $90\%$ of the total cluster mass has assembled. The horizontal dashed line denotes the main sequence lifetime of a $30\ {\rm M}_\sun$ star, which is the earliest time a star can deposit energy and momentum in a SN in our simulations.}
    \label{fig:t90t10}
\end{figure*}

Figure~\ref{fig:t90t10} shows the time between the formation of the first star in a cluster $t_0$ and the final time $t_{100\%}$, and the time it takes each cluster to go from $10\%$ to $90\%$ of its final cluster mass.
Note that $t_{100\%}-t_{0}$ is more sensitive to outliers and resolution since these points are often the result of a few star particles that formed at a different time compared to the bulk of the cluster members.
Without feedback, many clusters grow in mass over long periods, with many cluster masses limited only by the maximum age of $25$~Myr for stars considered by our cluster-finding algorithm.
The clusters sitting at the upper limit in formation time confirm that the overall mass function is not converged with the age limits applied in this case, due to continued cluster growth (see also Figure~\ref{fig:nofb_dndm}).
This is not the case for the remaining simulations, and as shown in Appendix~\ref{sec:fof_parameters}, the mass function has converged for all age limits $\geq25$~Myr.

In Figure~\ref{fig:t90t10} the lower limit to cluster age at each mass is roughly linearly proportional to the log of the cluster mass. 
Furthermore, in the simulations with SNe and SNe+winds, all clusters scatter around such a trend.
We note that the linear trend limiting the lowest mass cluster has a shallower slope in simulations with radiation, indicating that more rapid growth is necessary to grow to larger masses in these simulations.
When including radiation, no single trend is found between the formation time and the cluster mass, indicating a complexity in what determines the growth rate in this case.
Similarly, \citet{Li&Gnedin&Gnedin2018} found that the cluster growth rate cannot be described by a simple power law in agreement with predictions for the collapse of turbulent clouds \citep[][]{Murray&Chang2015,Murray+2017}.

Many clusters grow past the first possible time for a SN explosion.
A stochastic sampling of the IMF for SN explosions implies that not all clusters contain a $30\ {\rm M}_\sun$ progenitor, with less massive clusters being more sensitive to this stochasticity.
When cluster growth is unlimited, low-mass clusters can grow quickly, thereby eliminating the possibility of clusters with low mass and long formation time in Figure~\ref{fig:t90t10}.
In the case when pre-SNe feedback is active, cluster growth is suppressed by feedback from the time when the first star forms, without any regard for stochasticity. \footnote{In our model, feedback from radiation and winds are computed by IMF-weighted averages, implying identical injection for clusters of the same mass.
Therefore, our simulations do not treat the stochasticity inherent to the IMF for these sources.
To remedy this, future models should incorporate this stochasticity made possible by star-by-star calculations (see discussion in Section~\ref{sec:limitations}).}
This results in low-mass clusters with long formation times, highlighting that pre-SNe feedback is not always strong enough to disrupt the cloud and stop star formation entirely.
Many of these low-mass clusters form stars past the time of the first possible SN.

\section{Discussion}\label{sec:Discussion}

\subsection{Cluster formation efficiency}

\begin{figure}
    \centering
    \includegraphics[width=0.45\textwidth]{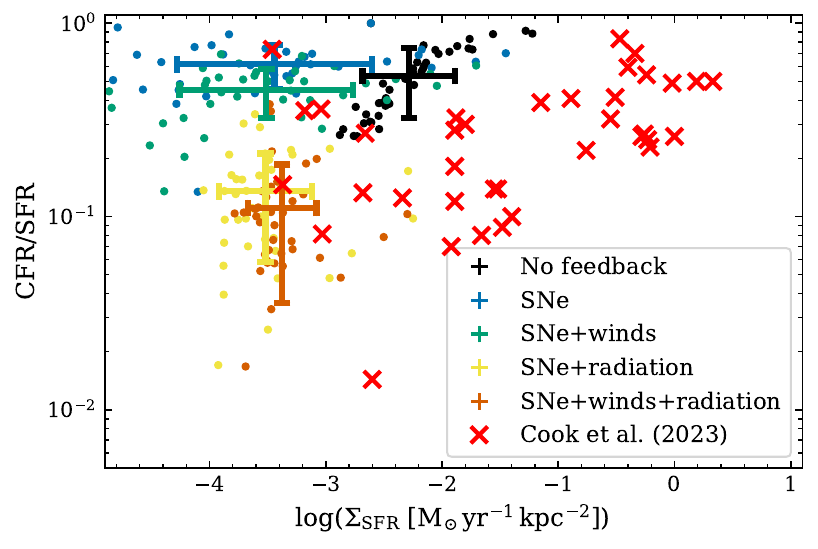}
    \caption{Cluster formation efficiency as a function of star formation rate surface density. Colored points show individual outputs, while the colored error bars show the standard deviation around the mean of all points for a given simulation. Red crosses show estimates from observed galaxies considering young (1--10~Myr old) star clusters. Observed data shown are compiled by \citet{Cook+2023} and include estimates from \citet{Goddard+2010,Adamo+2011,Annibali+2011,Pasquali+2011,Cook+2012,Adamo+2015,Lim&Lee2015,Hollyhead+2016,Chandar+2017,Messa+2018,Adamo+2020b}.}
    \label{fig:CFE_SSFR}
\end{figure}

We find that feedback introduced immediately after star formation (e.g., stellar winds and radiation) limits the formation of massive star clusters and significantly reduces the cluster formation efficiency.
In Figure~\ref{fig:CFE_SSFR}, we show the cluster formation efficiency of each output as a function of the star formation rate surface density $\Sigma_{\rm SFR}$.
There is notable scatter in the CFE for similar values in $\Sigma_{\rm SFR}$ in all simulations, as highlighted by the mean value and standard deviation shown with the error bars.
\citet{Lahen+2023} found a similar scatter in a simulation of a comparable galaxy, albeit with the majority of points sitting at higher efficiency. 
Note that \citet{Lahen+2023} defined clusters with a lower minimum mass ($100\ {\rm M}_\sun$) than us, which by definition may lead to a higher efficiency (our choice does not consider clusters with mass $<500\ {\rm M}_\sun$).
Furthermore, there is a discrepancy between the cluster ages ($10$~Myr versus our $25$~Myr), however, decreasing the star age limit to $10$~Myr would not affect our results in a measurable way (see Appendix~\ref{sec:fof_parameters}).
We find a wide range of CFE in multiple snapshots at similar $\Sigma_{\rm SFR}$ in our simulations with feedback. Furthermore, CFE has roughly the same mean value over almost three orders of magnitude in $\Sigma_{\rm SFR}$ in SNe and SNe+wind, although the scatter increase towards lower $\Sigma_{\rm SFR}$.
In the simulation without feedback, we do find a clear trend between different outputs in this space.
In this case, the trend has a clear evolution in time (later outputs toward the bottom left) and coincides with a reduced gas fraction over time (our simulations lack significant gas inflow), although causality remains unconfirmed (other factors likely play a role, e.g., morphological evolution).

The red crosses in Figure~\ref{fig:CFE_SSFR} show observational estimates from the literature \citep[][see also figure caption]{Cook+2023}, and indicate an absence of low CFE at higher $\Sigma_{\rm SFR}$.
Note that this is debated in the literature, in particular concerning the age limit of the observed clusters \citep[see, e.g.,][]{Chandar+2017,Chandar+2023}.
Furthermore, these estimates rely on different methods, although all values shown in Figure~\ref{fig:CFE_SSFR} consider only clusters in the age range 1--10~Myr \citep[see][for discussions about the different methods employed]{Adamo+2020a,Cook+2023}.
In our simulations, changing the age limit to $10$~Myr does not affect the CFE when including feedback (see Appendix \ref{sec:fof_parameters}) since the vast majority of star clusters have stopped growing by this time.
Since we only consider one dwarf galaxy in our simulations, we cannot address the question of high CFE and $\Sigma_{\rm SFR}$ at higher $\Sigma_{\rm SFR}$. 

Extrapolating the effect of feedback on star cluster formation to more massive galaxies (and thereby higher $\Sigma_{\rm SFR}$) from our models is not trivial, since many properties to which the cluster formation process is sensitive change. 
For example, more massive systems have higher metal content, which affects gas dynamics through cooling processes \citep[][]{Ferland+1998,Rosen&Bergman1995} and through the ability for feedback to couple dynamically to the gas, both for radiation \citep[via dust attenuation][]{Rosdahl&Teyssier2015}, and strength of stellar winds \citep[][]{Vink2011}.
This is complicated further by other properties that vary between galaxies, such as gas fraction and merger events, which have an impact on the dynamics of the galaxy and the population of clusters \citep[][]{Li+2005,Renaud+2019,Renaud+2021}. 

\subsection{Maximum cluster mass}

From Figure~\ref{fig:mass_function}, it is obvious from our models that increasing the number of active feedback mechanisms  reduces the maximum cluster mass.
An upper mass limit or a Schechter-like function with an exponential decline above a mass threshold is often reported in the literature \citep[see reviews by][]{PortegiesZwart&McMillan&Gieles2010,Krumholz&McKee&Bland-Hawthorn2019,Adamo+2020a}.
In the case of a truncation mass, several works report a value around $10^{5}\ {\rm M}_\sun$ for spiral galaxies \citep[][]{Gieles+2006,Bastian+2012,Adamo+2015,Messa+2018}, although notable exceptions are galaxies experiencing mergers \citep[$\sim10^{7}\ {\rm M}_\sun$,][]{Adamo+2020b} and the M31 galaxy \citep[$\lesssim10^{4}\ {\rm M}_\sun$,][]{Johnson+2017}.
On the other hand, \citet{Mok+2019} found many cases where high-mass truncation was not deemed significant when accounting for completeness and errors in the cluster mass estimates, instead reporting a pure power law for several local galaxies.
The qualitative shape of the mass functions are different between our model; SNe+winds+radiation is well described by a power law, while no feedback is neither described by a power law nor a Schechter function but instead approaches a log-normal.
Extracting a cluster mass function for dwarf galaxies accurately enough to determine the presence of a truncation mass is highly challenging due to the low number of clusters in any one galaxy, and likely requires stacking observations of several galaxies.
Nonetheless, we emphasize the importance of considering a range of feedback mechanisms (stellar winds, radiation, and SNe) when formulating a general theoretical framework for star cluster formation in different galaxies. 

\subsection{Maximum SN progenitor mass}\label{subsec:progenitor_mass}

\begin{figure}
    \centering
    \includegraphics[width=0.45\textwidth]{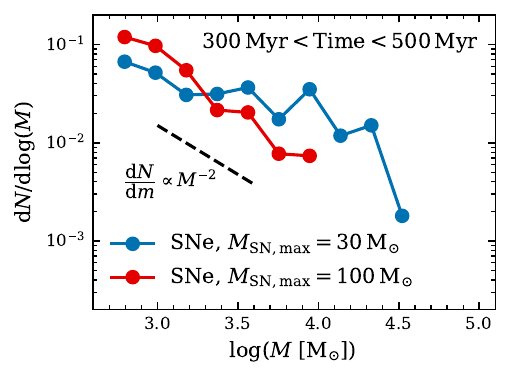}
    \caption{Cluster mass function (cf. Figure~\ref{fig:mass_function}) for the simulation with the fiducial SN feedback model compared to the mass function for a simulation where stars with mass up to $100\ {\rm M}_\sun$ can explode as core-collapse SNe (see text for details). Allowing more massive stars to explode as SNe implies an earlier onset of SN feedback due to the shorter time spent on the main sequence. Because the simulation with massive SN progenitors only progressed $500$~Myr, we limit the results to clusters formed between $300$ and $500$~Myr. Note that the earlier onset of SNe results in a mass function without an obvious plateau at lower masses.}
    \label{fig:dndm_earlySNe}
\end{figure}

In the case of SNe, we limit the injection of mass, momentum, and energy to stars within the mass range 8--30$\ {\rm M}_\sun$, assuming direct collapse for more massive stars.
The direct collapse to a black hole for $M>30\ {\rm M}_\sun$ is predicted by many theoretical models of the core-collapse process \citep[][]{OConnor&Ott2011,Ertl+2016,Sukhbold+2016}, although there is notable uncertainty, particularly with regards to explosion mechanism \citep[][]{Janka2012}.
Our choice of $30\ {\rm M}_\sun$ translates to a minimum delay of $\sim6$~Myr between the formation of a star particle and the first release of energy from SNe.
Furthermore, there are theoretical predictions that the most massive stars tend to form late during the build-up of the IMF \citep[][]{Haugbolle+2018,Grudic+2023}.
To investigate how our choice affects our conclusions, we present the mass function for a re-simulated SNe model with an SN mass threshold at $100\ {\rm M}_\sun$ (resulting in a $3.3$~Myr main sequence lifetime) in Figure~\ref{fig:dndm_earlySNe}.
In this simulation, we limit the explosions to energy, momentum, and mass (i.e., we do not inject enriched material to maintain a similar gas metallicity).
The new simulation ran for $500$~Myr. Therefore, we limit the comparison to this time for both simulations presented in Figure~\ref{fig:dndm_earlySNe}.
Similarly to the simulation with radiation and winds, earlier SNe from the more massive stars reduces the number of the most massive clusters, resulting in a power law-like mass function for the entire range of masses.
Note that changing the maximum mass for SN progenitors not only affects the timing of feedback onset but also the energy budget.
The sensitivity to choices with regard to uncertainties in SN modeling highlights that uncertainty remains, even for state-of-the-art galaxy models.

\subsection{Limitations of our model}\label{sec:limitations}

Our results are derived from numerical simulations of a single galaxy in an isolated environment. This restricts our conclusions to the properties of this specific galaxy, e.g., one initial gas fraction, gravitational background, and galactic dynamical timescale. 
Furthermore, this limits our range of the turbulence spectrum, Mach numbers, and cloud-to-inter-cloud density contrasts, which are suspected to be key ingredients in the assembly of the star cluster forming sites \citep[e.g.][]{Renaud+2019, Li+2022}.
Furthermore, our simulations rely on parameter choices that are debated in the literature (e.g., the star formation efficiency per free-fall time).
Therefore, we emphasize that our results should be interpreted as relative changes between our simulated models.

Although our model incorporates the most common mechanisms for stellar feedback, it still lacks several processes, e.g., cosmic ray physics, massive rapidly rotating stars (Wolf-Rayet stars), stellar multiples, exotic SNe (e.g., pair-instability SNe), and protostellar jets.
Of particular interest are assumptions about the stellar feedback mechanisms that affect the local gas dynamics soon after the star forms (see also Section~\ref{subsec:progenitor_mass}).
For the star particles, our model assumes a Kroupa IMF \citep[][]{Kroupa2001} with an upper mass limit of $100\ {\rm M}_\sun$ for calibrating the input for the radiative transfer, calculating the mass loss from stellar winds, and stochastically sampling progenitors for SNe events.

Although our model stochastically samples SNe as discrete events, winds, and radiation are calibrated to IMF averages.
This can affect both galactic scale properties \citep[][]{Smith2021}, as well as affect the final cluster properties \citep[][]{Lewis+2023}.
An ideal tool to study these effects are star-by-star models which have recently become common \citep[see, e.g.,][]{Hu+2016,Emerick+2018,Andersson+2020,Gutcke+2021a,Hirai+2021,Hislop+2021,Andersson+2023,Lahen+2023}, and should be applied to this problem in the near future \citep[see][specifically]{Hislop+2021,Lahen+2023}.

Our simulations are initialized from the same initial conditions; however, different feedback mechanisms are active from the start.
Since these mechanisms are of a fundamentally distinct nature, differences could accumulate over time and change the environment where the clusters form.
As an alternative, the simulations can be run with similar models to some point after which the different feedback mechanisms are applied.
This approach is better suited for studying the role different feedback mechanisms play in the cluster formation process in an identical environment.
However, a challenge with this approach is the inconsistency in the ISM structure set by a different stellar feedback cycle and thereby different turbulence.
Performing these types of tests is beyond the scope of this paper, but see forthcoming work by Ringdahl et al. (in preparation).
The cluster formation process in controlled environments can also be tested simulation isolating the environment from initialization on smaller scales \citep[see, e.g.,][]{Wall+2019,Wall+2020,Cournoyer-Cloutier+2021,Cournoyer-Cloutier+2023,Lewis+2023}.

Another limitation of our model is unresolved dynamical interactions between particles inside the star clusters because of the lack of star-by-star calculations and systems of multiple stars (e.g., binaries).
Furthermore, the variations in the tidal field around the clusters imposed by gas motion are only marginally resolved ($~4$~pc).
Such variations can affect the cluster evolution \citep{Renaud2018, Renaud2020}.
These are important for the dynamical properties of clusters and their evolution, and as such, we limit our study to young clusters and refrain from studying properties where the internal dynamics play a role (e.g., size and velocity dispersion).
Due to the computational demand of direct $N$-body calculations, studying the dynamical evolution of star clusters is typically limited to individual clusters in gas-free simulations \citep[see, e.g.,][]{Renaud+2011, Wang+2016}, or including gas until it is expelled from an individual cluster \citep[see, e.g.,][]{Wall+2019,Wall+2020,Ballone+2021,Cournoyer-Cloutier+2021,Grudic+2021,Guszejnov+2022,Cournoyer-Cloutier+2023}, although preliminary work on the topic was done by \citet{mamikonyan2017}.

\section{Summary}\label{sec:Summary}

Our work studies how stellar feedback affects the population of young star clusters.  To do this, we test different combinations of feedback in hydrodynamical simulations of isolated dwarf galaxies with an initial stellar mass of $10^7\ {\rm M}_\sun$. 
For the first time, we systematically compare combinations of different feedback sources to investigate the cluster initial mass function and CFE.
We find the following:
\begin{enumerate}
\item Introducing radiation changes the cluster formation efficiency dramatically (see Figures~\ref{fig:cfr} and \ref{fig:CFE_SSFR}). The total stellar mass in young clusters relative to the total stellar mass formed is $\sim0.6$ in simulations with only SNe or SNe+winds, while for simulations with additional radiation feedback, the same fraction is $\sim0.15$.
\item We find that systematically removing feedback mechanisms transforms the upper end of the initial cluster mass function (see Figure~\ref{fig:mass_function}). In the simulation with SNe+winds+radiation, the initial cluster mass function follows a power law with a slope around $-3$. As mechanisms resulting in feedback are removed, the power law shifts to larger masses with a plateau at lower masses. In the absence of feedback, the function resembles a log normal.
\item  The total number of low mass clusters ($M\la10^3\ {\rm M}_\sun$) is similar in all simulations (see Figure~\ref{fig:mass_histrogram}). Therefore, the cluster formation efficiency is determined by the abundance of more massive clusters.
\item The timing of feedback initiation is crucial to the formation of star clusters. Feedback with early onset after star formation, such as winds and radiation, prevents cluster growth and explains the lack of massive clusters as compared to simulations without feedback or only SNe. Figure~\ref{fig:t90t10} show how the cluster formation time varies between the models with different feedback mechanisms active.
\item We employ two tests to check how the cluster formation timescale affects the cluster mass function:
\begin{enumerate}
\item by placing progressively lower limits on the age of stars considered to be cluster members, we find that the initial cluster mass function in the simulation without feedback becomes progressively more similar to the simulations with earlier onset of feedback (see Figure~\ref{fig:nofb_dndm}); 
\item by letting massive (up to $100\ {\rm M}_\sun$) stars deposit mass, momentum, and energy as SNe (relaxing the assumption of direct collapse to black hole for stars with $M > 30\ {\rm M}_\sun$), the initial cluster mass function show only a power law slope (albeit with shallower slope compared to simulation with radiation) even in a simulation with only SNe (see Figure~\ref{fig:dndm_earlySNe}). Note that these more massive SN progenitors imply earlier explosions due to the shorter time spent on the main sequence.
\end{enumerate}
\end{enumerate}

In conclusion, the formation of star clusters is sensitive to the timing for the onset of stellar feedback. Therefore, feedback that starts more quickly or slowly results in different CFE and initial cluster mass functions. 
The immediate regulation of local star formation provided by radiation and stellar winds results in a power-law mass function for young star clusters.
Without these sources of pre-SN feedback, star clusters rapidly grow in mass, shifting the entire mass function toward higher masses (with a plateau at lower masses) and high cluster formation efficiencies. 

\begin{acknowledgements}
EA thanks participants of the 2023 conferences {\it Phases of Galactic Evolution as Traced by Stellar Populations and Star clusters} and {\it A Multi-wavelength View on Globular Clusters Near and Far: From JWST to the ELT} for discussions that benefited this work. EA thanks John Ringdahl and Elin Sandvik for insightful conversations. EA acknowledges resources from SNIC 2022/6-76 (storage) and LU 2022/2-15 (computing) for executing the simulations. EA and M-MML acknowledge support from US NSF grants AST18-15461 and AST23-07950.
FR and OA acknowledge support from the Knut and Alice Wallenberg Foundation. FR acknowledges the support provided by the University of Strasbourg Institute for Advanced Study (USIAS), within the French national programme Investment for the Future (Excellence Initiative) IdEx-Unistra.
\end{acknowledgements}

\bibliographystyle{aa}
\bibliography{ref}

\begin{appendix}
\section{Sensitivity to cluster definition}\label{sec:fof_parameters}
\begin{figure}
    \centering
    \includegraphics[width=0.45\textwidth]{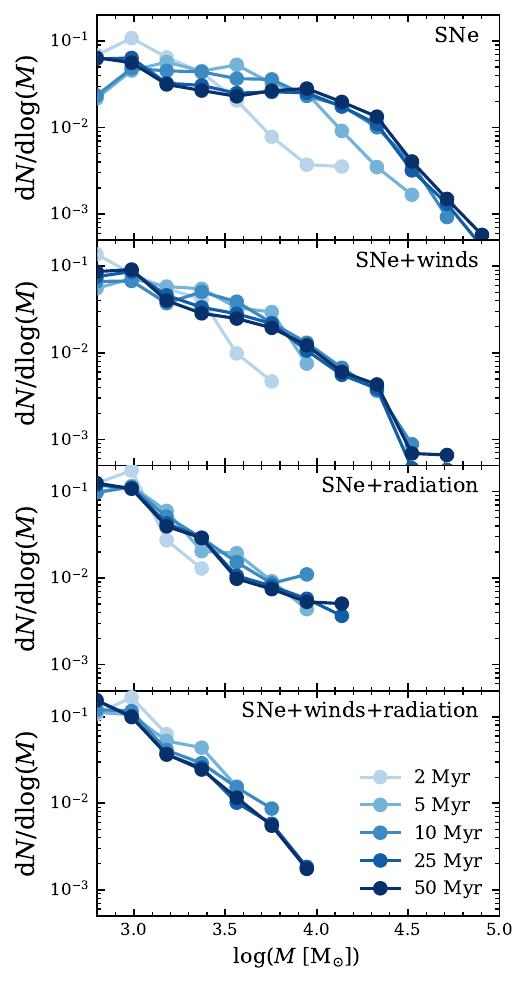}
    \caption{Cluster mass function for different stellar age limits. The simulation without feedback is shown in Figure~\ref{fig:nofb_dndm}.}
    \label{fig:dndm_tlim}
\end{figure}

\begin{figure}
    \centering
    \includegraphics[width=0.45\textwidth]{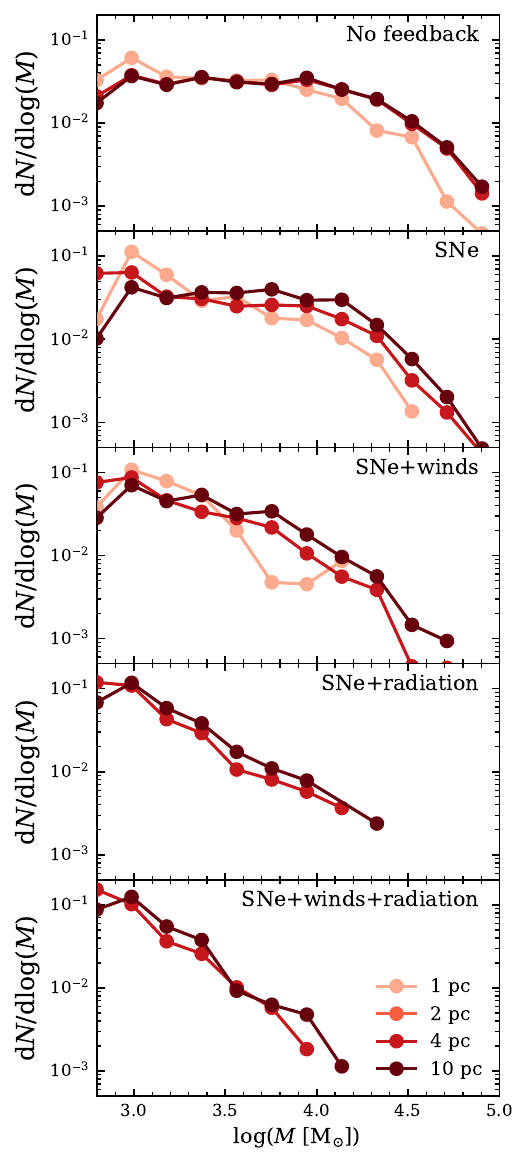}
    \caption{Cluster mass function for different choices of linking length. In the simulations with radiation, the smallest linking lengths result in no clusters.}
    \label{fig:dndm_llim}
\end{figure}

\begin{figure}
    \centering
    \includegraphics[width=0.45\textwidth]{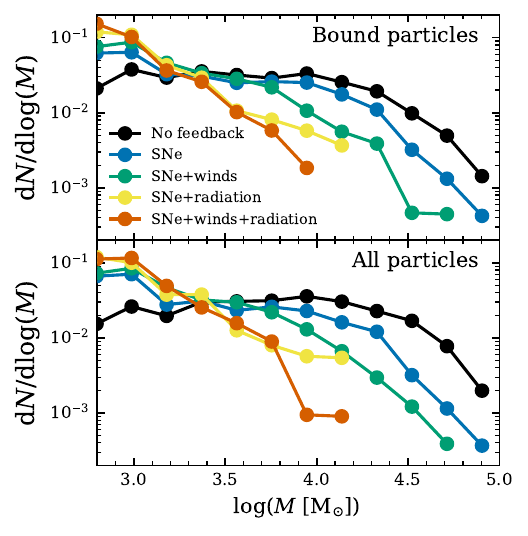}
    \caption{Mass function of all simulations. The top panel is identical to Figure~\ref{fig:mass_function}, while the bottom shows the result without applying the bound criterion.}
    \label{fig:dndm_bound}
\end{figure}

In this appendix, we show how our results depend on the assumptions used for selecting the clusters in our analysis.
Clusters are found by applying a friends-of-friends search to stars limited in age, connecting neighboring stars within a linking length.
The age limit was investigated for the simulation without feedback in Sections~\ref{sec:mass_function} and \ref{sec:formation_time}.
We present how the remaining simulations are affected by different limits to the stellar age in Figure~\ref{fig:dndm_tlim}.
While the slope of the mass function in simulations with radiation, which begins promptly, remains unaffected by this choice, the most massive cluster found converges for ages $>25$~Myr.
In the simulations lacking radiation, the shape of the mass function changes.
In fact, SNe and SNe+wind show a behavior similar to that found in Figure~\ref{fig:nofb_dndm}, but less pronounced.
Figure~\ref{fig:dndm_llim} shows how the choice of linking length affects the cluster mass function.
Our fiducial choice is $4$~pc, and our results are only marginally affected by increasing this linking length.
Note that the spatial resolution of our simulations is $3.6$~pc, and clusters smaller than this will suffer severely from numerical smoothing.

Finally, Figure~\ref{fig:dndm_bound} shows how the mass function is affected by removing unbound particles from the clusters. Note that the upper panel is identical to Figure~\ref{fig:mass_function}.
Surprisingly, most clusters are highly bound, and this criterion only marginally changes the mass distribution.
Note that, except for the cluster finding parameters, all figures in this section are produced in a way identical to Figure~\ref{fig:mass_function}.
\end{appendix}

\end{document}